\documentclass[twoside,twocolumn,english,aps,prx]{revtex4-1}

	\usepackage[usenames,dvipsnames]{xcolor}
	\usepackage{amsmath}
	\usepackage{amsfonts}
	\usepackage{amssymb}
	\usepackage[colorlinks=true,citecolor=blue,linkcolor=red]{hyperref}
	\usepackage{graphicx}
	\usepackage{bbold}					
	\usepackage[makeroom]{cancel}		
	\usepackage{multirow}				
	\usepackage[normalem]{ulem}        
	\usepackage{array}
	\usepackage{pdfpages}
	\makeatletter
	\AtBeginDocument{\let\LS@rot\@undefined}
	\makeatother
	
	\newcolumntype{x}[1]{>{\centering\let\newline\\\arraybackslash\hspace{0pt}}p{#1}}

	\DeclareMathAlphabet{\mathbbold}{U}{bbold}{m}{n}

	\DeclareMathOperator{\Tr}{Tr}
	
	\newcounter{subeqn} %
	\makeatletter
	\@addtoreset{subeqn}{equation}
	\makeatother

\definecolor{TB}{rgb}{0,0,0} 

\begin{document}
\title{Intrinsic dissipative Floquet superconductors beyond mean-field theory }

\author{Qinghong Yang$^{1}$}
\author{Zhesen Yang$^{2}$}
\author{Dong E. Liu$^{1,3,4}$}
\email{Corresponding to: dongeliu@mail.tsinghua.edu.cn}

\affiliation{$^{1}$State Key Laboratory of Low Dimensional Quantum Physics, Department of Physics, Tsinghua University, Beijing, 100084, China}
\affiliation{$^{2}$Kavli Institute for Theoretical Sciences, University of Chinese Academy of Sciences, Beijing 100190, China}
\affiliation{$^{3}$Beijing Academy of Quantum Information Sciences, Beijing 100193, China}
\affiliation{$^{4}$Frontier Science Center for Quantum Information, Beijing 100184, China}
\date{\today}

\begin{abstract}
We study the intrinsic superconductivity in a dissipative Floquet electronic system in the presence of attractive interactions. Based on the functional Keldysh theory beyond the mean-field treatment, we find that the system shows a time-periodic bosonic condensation and reaches an intrinsic dissipative Floquet superconducting (SC) phase. Due to the interplay between dissipations and periodic modulations, the Floquet SC gap becomes ``soft" and contains the diffusive fermionic modes with finite lifetimes. However, bosonic modes of the bosonic condensation are still propagating even in the presence of dissipations. 
\end{abstract}


\maketitle

\section{Introduction}\label{Intro}
Periodic driving schemes provide a simple way to study systems out of equilibrium~\cite{RevModPhys.89.011004,doi:10.1146,rudner2019floquet}, also known as Floquet engineering~\cite{Wyatt66,Galitskii69,Elesin71,inoue10,lindner11,kitagawaGF11,Dahlhaus11,jiang11,Kitagawa:2012aa,reynoso12,Liu13,IadecolaPRL13,Fregoso13,Iadecola14,FoaTorres14,Sedrakyan15,KitagawaNC12,RechtsmanNature,Struck13,FloquetClassification,FloquetPeriodicTable,BomantaraPRL18,BomantaraPRB18,PengYangPRB18,BauerPRB19,PhysRevLett.117.087402,PhysRevB.96.195303,McIverNP2020,SatoFloquet19,PhysRevB.101.201401}. An interesting example is the Floquet topological superconductor~\cite{jiang11,reynoso12, Liu13, FloquetClassification, BomantaraPRB18,dehghani2020optically}. Non-equilibrium superconductivity, including the enhancement of superconductivity due to non-equilibrium electrons~\cite{Eliashberg70,Galitskii73,Elesin73,AslamazovJETP82,Robertson09,GoldsteinPRB15,PhysRevX.10.031028,dehghani2020optically} or non-equilibrium phonons~\cite{Mankowsky14,PhysRevB.92.224517,PhysRevB.93.144506,PhysRevB.94.214504,PhysRevB.96.045125,PhysRevB.100.024513}, and dynamics of Cooper correlations due to time-dependent interactions~\cite{PhysRevLett.93.160401,PhysRevLett.96.230403,PhysRevLett.96.230404,PhysRevLett.115.257001,PhysRevA.97.013619,PhysRevA.98.053605,PhysRevB.99.174509,PhysRevB.101.054502,PhysRevB.98.214519,tindall2020dynamical}, has been widely studied and obtained great exciting results. In this paper, however, we want to understand Floquet structures and their behaviors in the presence of dissipations, which is unavoidable and a sensitive factor to Floquet engineering. Floquet superconductivity can be induced in two different routes: proximity-induced SC and intrinsic SC. The proximitized SC provides not only Cooper correlations but also strong dissipations, which could significantly change the behavior of Floquet systems~\cite{Liu,yang2020dissipative}. For the intrinsic case, SC is created due to the interaction instability near the Fermi surface or other strong interaction effects~\cite{Sigrist,Maiti_2013}.  Those interaction instabilities and related dissipations could be significantly modified by the periodic driving potential in Floquet engineering; and so, the Floquet treatment, i.e. Floquet theorem for quadratic BCS mean-field Hamiltonian, could be unreliable in the Floquet engineering. Therefore, a careful self-consistent treatment of all critical factors, i.e. periodical driving force, interaction instabilities and dissipations, should be considered for Floquet engineering. Based on those motivations, we focus on the questions: can the interaction cause fermion-to-boson transition in the dissipative Floquet systems; and how well is the Floquet BCS mean-field treatment in describing periodically driven intrinsic SC?

In this paper, we study Cooper instability for a realistic periodically driven electronic system with interactions and dissipations.  Based on the functional Keldysh field theory~\cite{Sieberer_2016,Kamenev}, we consider both the stationary point analysis and the Gaussian fluctuation, which is beyond the mean-field theory. We show that the system develops a periodic bosonic condensation, and reaches a dissipative  Floquet superconducting phase below a critical value $\gamma_c$ or $T_c$, where $\gamma$ is the system-bath coupling and $T$ is the temperature of the bath. The fermionic quasiparticle shows a ``soft" energy gap, acquires a finite lifetime and becomes diffusive. However, bosonic modes of the condensation are still propagating even in the presence of dissipations. In addition, we also find that the oscillation amplitude of the order parameter is a non-monotonic function of dissipation; and therefore, a certain finite dissipation will be helpful for the Floquet SC. 

\subsection*{Summary of the Treatment}
In order to avoid bringing confusions to readers and make our manuscript easy to read, we summarize the treatment we used in this manuscript. Our treatment is a generalization of the method used in the equilibrium superconducting case, to the dissipative Floquet system. The first step is to write down the Hamiltonian Eq.\eqref{eq:OriginalHamiltonian} of the composite system(electronic system with periodic driving $+$ normal metal bath) under consideration. We then apply a time-dependent unitary transformation shown in Eq.\eqref{eq:RotatingFrame} to obtain an equivalent time-independent electronic system (then we can directly use the knowledge of the static electronic system). We focus on the possibility of dissipative Floquet superconductivity in the presence of the bath. 
We further apply the functional Keldysh field theory~\cite{Kamenev} to obtain the action Eq.\eqref{eq:FermionicAction} from the Hamiltonian Eq.\eqref{eq:RotatingFrame}; and this treatment is a standard routine (see Appendix \ref{app:EffAction}). Next, we use the Hubbard-Stratonovich transformation to decouple the four-fermion attractive interaction by introducing an auxiliary {\em bosonic} field, and then integrate out fermionic degrees of freedom using Gaussian integrals, arriving at an effective bosonic field theory Eq.\eqref{eq:BosonicAction}(see Appendix \ref{app:EffAction}). Those are formulated in Sec.\ref{IDFS}. After that, based on the effective bosonic theory, we do the stationary point analysis in Sec.\ref{Sec:SPA} and the Gaussian fluctuation approximation, which is beyond the mean-field theory, in Sec.\ref{CI}. Generalized Feymann diagram rules are developed in Sec.\ref{CI} to facilitate the analysis.


\section{Intrinsic Dissipative Floquet SC}\label{IDFS}
We consider a single-band electronic system with a time-periodic chemical potential and attractive interactions, coupling to a normal fermionic bath. The Hamiltonian of the whole system can be written as 
\begin{equation}
\begin{split}
H(t)&=H_D+H_{int}+H_T+H_B,\\
\label{eq:OriginalHamiltonian}
\end{split}
\end{equation}
where $H_D=\sum_{\mathbf{k}\sigma}\left[\epsilon_{\mathbf{k}}-\mu_0-\mu(t)\right]c^{\dagger}_{\mathbf{k}\sigma}c_{\mathbf{k}\sigma}$ describes the non-interacting electronic system with the time-periodic chemical potential
 $\mu_0+\mu(t)$, where $\mu(t)=-K\cos(\Omega t)$ with $\Omega$ the driving frequency, and $c^{\dagger}(c)$ is the fermionic creation(annihilation) operator. $H_{int}=-g\sum_{\mathbf{q}\mathbf{k}_{1}\mathbf{k}_{2}}c_{\mathbf{k}_{1}+\mathbf{q}\uparrow}^{\dagger}c_{\mathbf{k}_{2}-\mathbf{q}\downarrow}^{\dagger}c_{\mathbf{k}_{2}\downarrow}c_{\mathbf{k}_{1}\uparrow}$  with $g>0$ describes the attractive interaction. Here, $H_B=\sum_{\mathbf{q}\sigma}\epsilon_{\mathbf{q}}a^{\dagger}_{\mathbf{q}\sigma}a_{\mathbf{q}\sigma}$ is the fermionic bath Hamiltonian, which provides dissipations; and $a^{\dagger}(a)$ is the creation(annihilation) operator of the bath. Such a bath is necessary for a driving interacting system to avoid the featureless infinite-temperature state~\cite{PONTE2015196} and thermalize to a non-trivial phase~\cite{rudner2019floquet}. One can imagine either an unavoidable dissipation resources or a large engineered equilibrium system weakly coupled to the small driven part; and the bath is in equilibrium state with temperature $T$. The system-bath coupling term can be written as $H_T=W\sum_{\mathbf{k}\mathbf{q}\sigma}\left(c^{\dagger}_{\mathbf{k}\sigma}a_{\mathbf{q}\sigma}+h.c.\right)$ with $W$ being the coupling strength.

For a periodically driven system, it is convenient to consider a rotating frame~\cite{Liu,PhysRevLett.106.220402,PhysRevB.89.115425,PhysRevB.94.214504} by using of a time-dependent unitary transformation $U_F=\operatorname{exp}(-if(t)\sum_{\mathbf{k}\sigma}c^{\dagger}_{\mathbf{k}\sigma}c_{\mathbf{k}\sigma})$ with $df/dt=-\mu(t)$, which results in an equivalent system  in the rotating frame
\begin{equation}\begin{split}
H_F(t)&=U_F^{\dagger}(H(t)-i\partial_t)U_F\\
&=\sum_{\mathbf{k}\sigma}(\epsilon_{\mathbf{k}}-\mu_0)c^{\dagger}_{\mathbf{k}\sigma}c_{\mathbf{k}\sigma}+H_{int}\\
&\quad+W\sum_{\mathbf{k}\mathbf{q}\sigma}\left(e^{if(t)}c^{\dagger}_{\mathbf{k}\sigma}a_{\mathbf{q}\sigma}+h.c.\right)+H_B.
\label{eq:RotatingFrame}
\end{split}\end{equation}
In the absence of the bath, that is $W\rightarrow 0$, we reach a time-independent system with equilibrium superconductivity, because the interaction term fully commutes with our periodic driven chemical potential term (thus commutes with $U_F$). Note that the original electron-phonon interaction also commutes with the periodic driven term, thus the phonon-induced attractive interaction is unchanged under the periodic driving. In those cases, without coupling to a bath, the periodic driving is trivial, and can be removed using a time-dependent unitary transformation. It is also our purpose to consider such a simple model that the periodical driving potential cannot simply spoil the formation of the SC correlation for the analysis of Floquet SC. In the presence of the bath with finite $W$, the superconductivity can be modified by the time-dependent system-bath coupling as shown in Eq.\eqref{eq:RotatingFrame}. 

Then, the standard procedure leads to the total action of the system in the closed time contour~\cite{Kamenev}:
\begin{equation}\label{eq:FermionicAction}
\begin{aligned}
S &=\int_cdt\int_cdt^{\prime}\sum_{\mathbf{k}}\vec{\Psi}^{\dagger}_{s\mathbf{k}}(t)\hat{Q}_{s0\mathbf{k}}^{-1}(t-t^{\prime})\vec{\Psi}_{s\mathbf{k}}(t^{\prime})+S_{int}\\
&\quad+\int_cdt\int_cdt^{\prime}\sum_{\mathbf{q}}\vec{\Psi}^{\dagger}_{b\mathbf{q}}(t)\hat{Q}_{b0\mathbf{q}}^{-1}(t-t^{\prime})\vec{\Psi}_{b\mathbf{q}}(t^{\prime})\\ 
&\quad+\int_cdt\sum_{\mathbf{{kq}}}\left[\vec{\Psi}^{\dagger}_{s\mathbf{k}}(t)\hat{M}(t)\vec{\Psi}_{b\mathbf{q}}(t)+h.c.\right],
\end{aligned}
\end{equation}
where $\vec{\Psi}_{s\mathbf{k}}=[\psi_{\mathbf{k}\uparrow},\bar{\psi}_{-\mathbf{k}\downarrow}]^t$, $\vec{\Psi}_{b\mathbf{q}}=[\phi_{\mathbf{q}\uparrow},\bar{\phi}_{-\mathbf{q}\downarrow}]^t$, $\hat{Q}_{s0\mathbf{k}}$ and $\hat{Q}_{b0\mathbf{q}}$ are free-fermion Green's functions in Nambu space,
\begin{equation*}
\hat{M}(t)=\left[
\begin{array}{ccc}
We^{if(t)} & 0\\
0 & -We^{-if(t)}
\end{array}
\right],
\end{equation*}
and in $S_{int}=g\int_cdt\sum_{\mathbf{k}\mathbf{k}^{\prime}}\bar{\psi}_{\mathbf{k}\uparrow}\bar{\psi}_{-\mathbf{k}\downarrow}\psi_{-\mathbf{k}^{\prime}\downarrow}\psi_{\mathbf{k}^{\prime}\uparrow}$, we only count in terms describing the interaction between electrons with opposite momenta and spins. 

In order to study the non-equilibrium steady state in such a system, one can first integrate out the bath's degrees of freedom with a Keldysh functional integral formalism~\cite{Liu,Kamenev} to obtain an equivalent effective fermionic action:
\begin{equation}
S^{\prime}=\int_cdt\int_cdt^{\prime}\sum_{\mathbf{k}}\vec{\Psi}^{\dagger}_{s\mathbf{k}}(t)\hat{Q}_{\mathbf{k}}^{-1}(t,t^{\prime})\vec{\Psi}_{s\mathbf{k}}(t^{\prime})+S_{int},
\end{equation}
where $\hat{Q}_{\mathbf{k}}(t,t^{\prime})$ is the dressed Green's function giving by the Dyson's equation
\begin{equation}\label{Eq:DE}
\begin{aligned}
\hat{Q}_{\mathbf{k}}(t,t^{\prime})&=\hat{Q}_{s0\mathbf{k}}(t-t^{\prime})\\
&\quad+\int_{-\infty}^{+\infty}dt_1dt_2\hat{Q}_{s0\mathbf{k}}(t-t_1)\hat{\Sigma}_{\mathbf{k}}(t_1,t_2)\hat{Q}_{\mathbf{k}}(t_2,t^{\prime}),
\end{aligned}
\end{equation}
with the self-energy from the bath being $\hat{\Sigma}_{\mathbf{k}}(t_1,t_2)=\sum_{\mathbf{q}}\hat{M}(t_1)\hat{Q}_{b0\mathbf{q}}(t_1-t_2)\hat{M}^{\dagger}(t_2)$.
These Green's functions can be derived analytically through perturbative expansions in the small parameter $\kappa=K/\Omega$(the driving potential reads as $\mu(t)=-K\cos(\Omega t)$)(see Appendix \ref{app:EffAction} for details).

For the four-fermion interaction, it is common to decouple them through the Hubbard-Stratonovich transformation~\cite{Altland,Kamenev}. Such a procedure will introduce an auxiliary bosonic field, denoted as $\Delta$ here. Applying the Keldysh transformation for the bosonic fields and the Keldysh-Lakin-Ovchinnikov transformation for the fermionic fields~\cite{Kamenev}, one turns the effective fermionic action---$S^{\prime}$ into the Keldysh$\otimes$Nambu space. Note that now $S^{\prime}$ is in a quadratic form with respect to the fermionic degrees of freedom, thus one can also integrate out fermionic fields and arrives at the effective bosonic action:
\begin{equation}\label{eq:BosonicAction}
\begin{split}
S_{\operatorname{eff}}=-\frac{1}{g}\int dt\bar{\Delta}^{\alpha}\hat{\sigma}_{1}^{\alpha\beta}\Delta^{\beta}(t)-i\Tr\ln\left[1-\breve{Q}_{\mathbf{k}}(\hat{\gamma}^{\bar{\alpha}}\otimes\hat{\Delta}^{\alpha})\right],
\end{split}
\end{equation}
where we have restricted the pairing between electrons with opposite momenta, thus $\Delta^{\alpha}$ here denotes the zero-momentum bosonic field; $\hat{\Delta}^{\alpha}=\left(1/\sqrt{2}\right)\left(\Delta^{\alpha}\hat{\tau}_{+}+\bar{\Delta}^{\alpha}\hat{\tau}_{-}\right)$ with $\alpha\in \{cl,q\}$ introduced by the Keldysh rotation; $\hat{\gamma}^{\bar{\alpha}}=\hat{\gamma}^{\alpha}\hat{\sigma}_1$, $\hat{\gamma}^{cl}=\hat{\sigma}_0$, $\hat{\gamma}^{q}=\hat{\sigma}_1$, and $\hat{\tau}_{\pm}=\hat{\sigma}_1\pm i\hat{\sigma}_{2}$, with $\hat{\sigma}_{\mu}(\mu=0,1,2,3)$ being the four Pauli matrices. One should not confuse the dissipation strength $\gamma$ with matrices $\hat{\gamma}^{\alpha}$ in the Keldysh space and we use the hat symbol for $2\times2$ matrices acting in either Nambu or Keldysh spaces and the check symbol to denote $4\times4$ matrices acting in the Keldysh$\otimes$Nambu space.

In the non-equilibrium case, one will often encounter two-time functions, like $\breve{Q}_{\mathbf{k}}(t,t^{\prime})$. Here, we show how to transform them into the Floquet representation~\cite{Liu} widely used in our discussion. 

Due to the periodic driving, those two-time functions will possess the discrete time-translational symmetry $Q(t,t^{\prime})=Q(t+\tau,t^{\prime}+\tau)$, where $\tau$ is the period. One can introduce two new variables $s\equiv t$ and $u\equiv t-t^{\prime}$ and define the new function $Q(t,t^{\prime})\rightarrow Q(s,u)$ which satisfies $Q(s+\tau,u)=Q(s,u)$ for all $u$. Thus, one can perform the Fourier transformation for $u$ and the Fourier series expansion for $s$:
\begin{equation}\label{Eq:FT}
\begin{aligned}
Q(s,\omega)&=\int_{-\infty}^{+\infty}du\, e^{i\omega u}\,Q(s,u),\\
Q(n,\omega)&=\frac{1}{\tau}\int_0^{\tau}ds\, e^{-in\Omega s}\,Q(s,\omega).
\end{aligned}
\end{equation}
Turning $Q(n,\omega)$ into a matrix form, known as the {\em Floquet structure}, one has
\begin{equation}
\underline{Q}=\begin{bmatrix}
\cdots \\ 
 &Q(0,\omega+\Omega) & Q(1,\omega) & Q(2,\omega-\Omega) & \\
 &Q(-1,\omega+\Omega) & Q(0,\omega) & Q(1,\omega-\Omega) & \\
 &Q(-2,\omega+\Omega) & Q(-1,\omega) & Q(0,\omega-\Omega) & \\
 & & & & \cdots
\end{bmatrix}\label{Eq:FS}.
\end{equation}

\section{Stationary Point Analysis}\label{Sec:SPA}
In equilibrium, the stationary point analysis of the effective bosonic action is just the mean-field theory and provides the gap equation. Out of equilibrium, we don't know what exactly the mean field is. However, we can always consider the stationary point, around which the action can be expanded perturbatively. A variation of the bosonic action Eq.\eqref{eq:BosonicAction} with respect to $\bar{\Delta}^{cl}(t)$ generates the stationary point equation:
\begin{equation}
\begin{split}
\Delta^q(t)&=i\frac{g}{\sqrt{2}}\Tr\left[\left(\hat{\gamma}^q\otimes\hat{\tau}_-\right)\sum_{\mathbf{k}}\breve{Q}_{\mathbf{k},\Delta}\right],\\
\end{split}\label{eq:GapEq}
\end{equation}
where in the time domain,
\begin{equation}
\begin{split}
\breve{Q}_{\mathbf{k},\Delta}^{-1}(t,t^\prime)=\breve{Q}_{\mathbf{k}}^{-1}(t,t^\prime)-\hat{\gamma}^{\bar{\alpha}}\otimes\hat{\Delta}^{\alpha}(t)\delta(t-t^\prime),
\end{split}
\end{equation}
and $\breve{Q}_{\mathbf{k}}(t,t^{\prime})$ is the non-interacting fermionic Green's function dressed by the self-energy of the bath in the Keldysh$\otimes$Nambu space as mentioned before. We want to ask: 1) if we have a stationary point solution for $\Delta^q(t)$; 2) if the solution describes the SC order parameter or "gap" of the dissipative Floquet superconductor. We will address the two questions below. 

\begin{figure}[t]
	\centerline{\includegraphics[height=6.3cm]{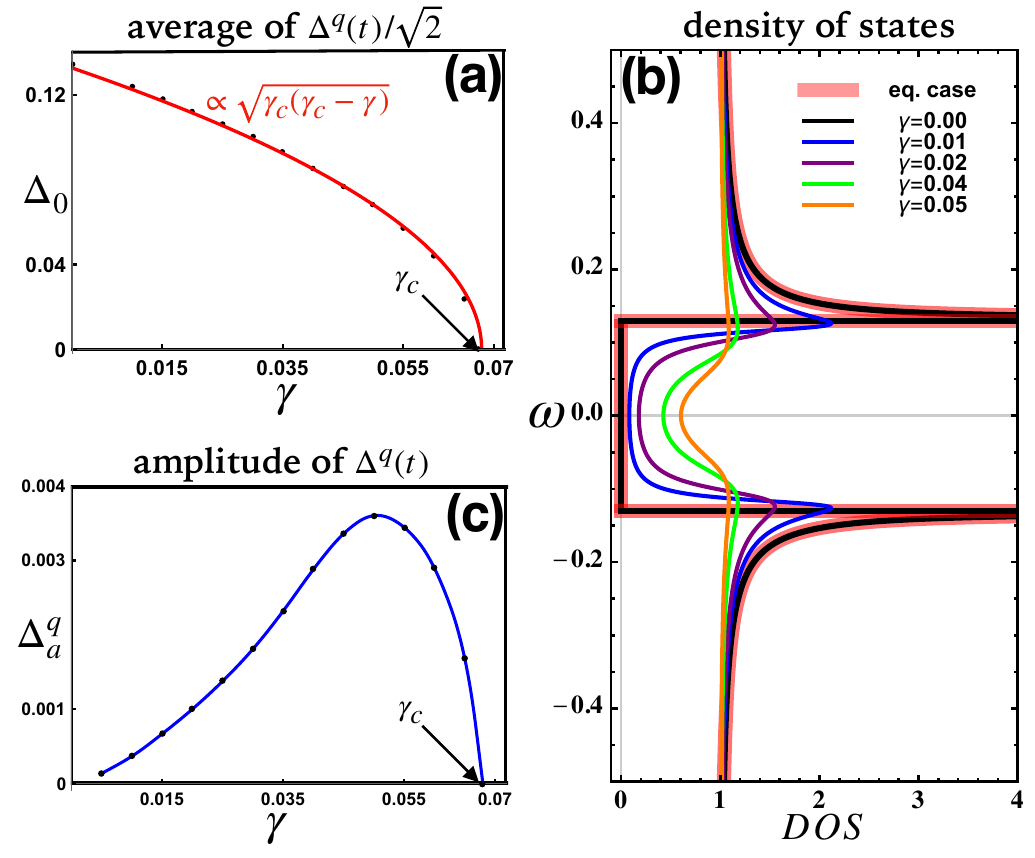}}
	\caption{These three figures are plotted under conditions: $\omega_D/\Omega=10,\Omega=100K$, $T=0K$, $g\rho_F=0.2$ and $\kappa=0.4$. Here $\omega$, $\Delta_0$, $\gamma$, and the amplitude of $\Delta^q(t)$ are all rescaled by being divided by $\Omega$. (a) The tendency of $\Delta_0$, i.e. the soft gap with the effective temperature $\gamma$. It scales as $\propto\!\!\sqrt{\gamma_c(\gamma_c-\gamma)}$, where $\gamma_c$ is the transition temperature. This scaling suggests that the critical value is $1/2$ ;(b) The DOS of quasiparticles in our intrinsic dissipative Floquet superconductors; (c) The amplitude of the order parameter $\Delta^q(t)$ for different $\gamma$'s. Note that the amplitude of $\Delta^q(t)$ is non-monotonic, as the amplitude is proportional to $\gamma\Delta_0$ and $\Delta_0$ changes in the opposite direction with the changing of $\gamma$.
		\label{Fig:DOS&Deltat}}
\end{figure}

As the system is periodic in the time domain, it is natural to assume that $\breve{Q}_{\mathbf{k},\Delta}(t,t^\prime)=\breve{Q}_{\mathbf{k},\Delta}(t+\tau,t^\prime+\tau)$, where $\tau$ is the period. Indeed, the validity of this {\em ansatz} will be confirmed later. Then, in the stationary point, the dominated field $\Delta^q(t)$ is also periodic in time. The Fourier transformation and the Fourier series expansion shown in Eq.\eqref{Eq:FT} lead to 
\begin{equation}
\begin{split}\label{dnq}
\Delta_{n}^{q}&=i\frac{g}{\sqrt{2}}\int\frac{d\omega}{2\pi}\Tr\left[(\hat{\gamma}^{q}\otimes\hat{\tau}_{-})\sum_{\mathbf{k}}\breve{Q}_{\mathbf{k},\Delta}\left(n,\omega\right)\right],
\end{split}
\end{equation}
where $\breve{Q}_{\mathbf{k},\Delta}(n,\omega)$ is one matrix element of the Floquet matrix---$\underline{\breve{Q}_{\mathbf{k},\Delta}}$ which has an infinite-dimension structure, and can be expressed as 
\begin{equation}\begin{split}\label{qkdn}
\underline{\breve{Q}_{\mathbf{k},\Delta}^{-1}}&=\underline{\breve{Q}_{\mathbf{k}}^{-1}}\\
&\quad-\frac{1}{\sqrt{2}}\left[\underline{\Delta^{\alpha(n)}}\otimes\left(\hat{\gamma}^{\bar{\alpha}}\otimes\hat{\tau}_{+}\right)+\underline{\bar{\Delta}^{\alpha(n)}}\otimes\left(\hat{\gamma}^{\bar{\alpha}}\otimes\hat{\tau}_{-}\right)\right],
\end{split}\end{equation}
where the repeated index $\alpha$ denotes the summation over $\alpha\in\{cl,q\}$, a general object $\underline{X}$ has the so-called Floquet structure shown in Eq.\eqref{Eq:FS}, and the superscript $(n)$ in $\underline{\Delta^{\alpha(n)}}$(or $\underline{\bar{\Delta}^{\alpha(n)}}$) with $n\in\mathbb{N}$ stands for the $i$th order in $\kappa$, e.g.,
\begin{equation}
\underline{\Delta^{\alpha(1)}}=\begin{bmatrix}
\cdots \\ 
 &0 & \Delta^{\alpha}_1 & 0 & \\
 &\Delta^{\alpha}_{-1} & 0 & \Delta^{\alpha}_1 & \\
 &0 & \Delta^{\alpha}_{-1} & 0 & \\
 & & & & \cdots
\end{bmatrix},
\end{equation}
where $\Delta^{\alpha}_n$ comes from the Fourier series expansion of $\Delta^{\alpha}(t)$.

Now, let's solve the stationary point equation Eq.\eqref{eq:GapEq} for the solution $\Delta^q(t)=\sum_n{\Delta_{n}^{q} e^{i n \Omega t}}$ with $n\in \mathbb{Z}$. Here, we consider small  $\kappa=K/\Omega$, and only keep terms up to leading order $\mathcal{O}(\kappa)$. The details are shown in Appendix \ref{App:SSPE}.

For the zero harmonic---$\Delta_0^q$, the stationary equation is exactly the gap equation in the equilibrium superconducting case when $\gamma\rightarrow 0^+$, where $\gamma=\pi\rho_FW^2$ is the dissipation strength provided by the self-energy from the bath, with $\rho_F$ being the density of states(DOS) in the vicinity of Fermi surface of the fermionic bath.
For finite $\gamma$,  we reach the following equation: 
\begin{equation}\label{ge}
1=g\rho_{F}\int^{\omega_{D}}_{\Delta_0}d\omega\frac{\cos\frac{\theta}{2}\tanh\frac{\omega}{2T}}{\left[\left(\omega^{2}-\gamma^{2}-\Delta_0^{2}\right)^{2}+4\omega^{2}\gamma^{2}\right]^{1/4}},
\end{equation}
where we define $\Delta_0\equiv(1/\sqrt{2})\Delta^q_0$ and $\tan\theta=2\omega\gamma/(\omega^{2}-\gamma^{2}-\Delta_0^{2})$, and $\omega_D$ is the Debye frequency. The numerical result for $\Delta_0$ is shown in Fig.\ref{Fig:DOS&Deltat}(a) indicating a special scaling behavior $\Delta_0\propto\sqrt{\gamma_c(\gamma_c-\gamma)}$(fitting from data points) and approaches zero at $\gamma_c$ , which has the same feature as its equilibrium counterpart. Thus, $\gamma$ could be regarded as the effective temperature, and the critical value $\gamma_c$ as the transition temperature(we will analytically derive it later). DOS of quasiparticles in the rotating frame can be derived from the following expression\begin{equation}\begin{aligned}
\nu(\omega)=-\frac{1}{\pi} \operatorname{Im} \Tr\left[\breve{Q}_{\Delta}^{R}(0, \omega)\right],\label{DOS}
\end{aligned}\end{equation}
 based on the $n=0$ component---$\breve{Q}_{\Delta}^R(0,\omega)$ of the quasi-classical Green's function \cite{RevModPhys.58.323,ch2003introduction}, which is a matrix element of $\underline{\breve{Q}^R_{\Delta}}$ in the Floquet space. As shown in Fig. \ref{Fig:DOS&Deltat}(a)(b), DOS of quasiparticles in the vicinity of the superconducting gap damps as the dissipation increases. Moreover, there still exists fermionic modes within the gap $\omega\in [-\Delta_0,+\Delta_0]$ when $\gamma$ is finite, and therefore, we consider them as "soft" gaps. The Green's function of quasiparticles---$\breve{Q}_{\mathbf{k},\Delta}^R(0,\omega)$ shows a finite imaginary part in the denominator due to the interplay between the dissipation and the periodical modulation; and therefore, the quasiparticle acquires a finite lifetime and becomes diffusive.

For our Floquet system, we should regard the time-resolved field $\Delta^q(t)$ obtained from the full stationary equation as the SC order parameter, which includes higher harmonics. Up to leading orders of $\kappa$, only lowest two harmonic components $\Delta^q_{0}$ and $\Delta^q_{\pm1}$ are relevant. In Fig.$\,$\ref{Fig:DOS&Deltat}(c), we numerically plot the amplitude of the SC order parameter as a function of the dissipation strength $\gamma$, defined as $\Delta_a^q=\max_{t\in[0,\tau]}(\Delta_{+1}^qe^{i\Omega t}+\Delta_{-1}^qe^{-i\Omega t})$. One can observe that $\Delta_a^q$ is a non-monotonic function of $\gamma$: as increasing $\gamma$, $\Delta_a^q$ first increases, and then starts to decrease after crossing a turning point. For larger $\gamma$, dissipations suppress the periodic modulation of the Floquet superconductor, and both the average value and the oscillation amplitude will drop to zero after $\gamma_c$. We also note that the amplitude $\Delta_a^q$ also approaches to zero in the limit $\gamma\rightarrow 0$. This comes from the fact that if $\gamma=0$, the system-bath coupling disappears. Therefore, in the rotating frame as shown in Eq.\eqref{eq:RotatingFrame}, the Floquet Hamiltonian reduces to its equilibrium counterpart, and the bosonic condensation is no longer periodic in time. Note that at a certain value of $\gamma$, the amplitude of the order parameter $\Delta^q(t)$ reaches a maximum. This comes from the competition of two effects resulted from the dissipation strength. On one hand, in the rotating frame, the periodicity of the order parameter comes from the system-bath coupling. Thus, we need $\gamma$ to be large in order that the fermionic system has a clear periodicity. On the other hand, we need $\gamma$ to be small in order that it will not kill the order parameter. Thus, the competition of these two effects results in the maximum value. From this physical interpretation, one can think that at this certain value of $\gamma$, the periodicity of the order parameter is most clear and stable, which makes it easier to detect this periodicity experimentally.

In fact, this stationary point analysis can be regarded as a Floquet BCS mean-field treatment in our self-consistent functional formalism. To obtain a more comprehensive understanding, let's go beyond this mean-field treatment.


\section{Formation of Periodic Bosonic Condensation}\label{CI}
From the previous discussion, we find that the bosonic field in the stationary point has a non-zero value when $\gamma$ is below a critical value $\gamma_c$. One can think that the non-zero bosonic field results from the condensation of Cooper pairs, just as the equilibrium SC case. The fact that some bosonic condensation exists in a pure fermionic system implies a fermion-to-boson phase transition\footnote{The reason that we use this phrase: Originally our system is a fermionic system, but below $\gamma_c$, we find that the {\em bosonic field $\Delta$} exists a non-zero value in the stationary point. This implies that the system now exists something kind of like bosons and this must come from the fermionic degrees of freedom in the fermionic system. Therefore, we designate this phenomenon as the fermion-to-boson transition}. We now discuss how this bosonic condensation is formed in the dissipative Floquet systems.

In the vicinity of $\gamma_c$, $\Delta^{cl(q)}(t)$ is small, which means one can expand $S_{\operatorname{eff}}$ around the critical point. Thus, we expand the $\Tr\ln$-term in $S_{\operatorname{eff}}$ in powers of $\Delta^{cl(q)}$, and simply keep terms up to the second order in $\Delta^{cl(q)}$. This can be easily achieved by using of the series expansion 
$\ln(1+x)=-\sum_{n=1}^{\infty}\frac{1}{n}(-x)^n$. Then applying transformations shown in Eq.\eqref{Eq:FT}, one can obtain the Gaussian action in the frequency space
\begin{equation}\begin{split}\label{Eq:SOAoBEA}
 S^{(2)}=&\int_0^{\Omega}\frac{d\omega}{2\pi}\;\underline{\bar{\vec{\Delta}}(\omega)}\left[-\frac{1}{g}\underline{\mathbb{1}}\otimes\hat{\sigma}_1+\sum_{\mathbf{k}}\underline{\hat{\Gamma}_{\mathbf{k}}}\right]\underline{\vec{\Delta}(\omega)},
\end{split}\end{equation}
where to avoid redundancy, we have restricted the integration range in $\left[0,\Omega\right]$, known as the first Floquet-Brillouin zone(FBZ)~\cite{PhysRevLett.95.260404,yang2020dissipative}, and the vector is defined as $\vec{\Delta}=\left[\Delta^{cl}, \Delta^{q}\right]^t$, and $\bar{\vec{\Delta}}=\left[\bar{\Delta}^{cl}, \bar{\Delta}^{q}\right]$. The Floquet matrix structure denoted by the underline has already defined in Eq.\eqref{Eq:FS}. Note that the identity in the Floquet space---$\underline{\mathbb{1}}$ only has diagonal elements and all of them are $\hat{\gamma}^{cl}$. 
The matrix elemets $\hat{\Gamma}_{\mathbf{k}}(n,\omega)$ of $\underline{\hat{\Gamma}_{\mathbf{k}}}$ comes from the Fourier transformation and the Fourier series expansion of $\hat{\Gamma}_{\mathbf{k}}^{\alpha\beta}(t,t^{\prime})$, which is defined as 
\begin{equation}\begin{split}
\hat{\Gamma}^{\alpha\beta}_{\mathbf{k}}(t,t^{\prime})&=\frac{i}{2}\Tr\left[\breve{Q}_{\mathbf{k}p}(t,t^{\prime})\hat{\gamma}^{\bar{\alpha}}\breve{Q}_{-\mathbf{k}h}(t^{\prime},t)\hat{\gamma}^{\bar{\beta}}\right],\\
\end{split}\end{equation}
with subscript $p$ and $h$ denoting particle and hole, respectively. 


Then, the Green's function of the bosonic field  in the Floquet space can be defined as 
\begin{equation}\begin{split}\label{Eq:BGF}
\quad\underline{\hat{D}}&=-i\langle\underline{\Phi(\omega)}\;\underline{\bar{\Phi}(\omega)}\rangle\\
&\sim-i\langle\begin{bmatrix}\vec{\Delta}\left(\omega+\Omega\right)\\
\vec{\Delta}\left(\omega\right)\\
\vec{\Delta}\left(\omega-\Omega\right)
\end{bmatrix}
\begin{bmatrix}\vec{\bar{\Delta}}\left(\omega+\Omega\right) & \vec{\bar{\Delta}}\left(\omega\right) & \vec{\bar{\Delta}}\left(\omega-\Omega\right)
\end{bmatrix}\rangle,
\end{split}\end{equation}
where $\langle\cdot\rangle$ describes the average with respect to the weight $\operatorname{exp}({iS^{(2)}})$. The function is defined in the Floquet basis $\underline{\Phi(\omega)}=[\cdots, \vec{\Delta}^t\left(\omega+\Omega\right), \vec{\Delta}^t\left(\omega\right), \vec{\Delta}^t\left(\omega-\Omega\right), \cdots]^t$, and
$\underline{\bar{\Phi}(\omega)}=[\cdots, \bar{\vec{\Delta}}\left(\omega+\Omega\right),  \bar{\vec{\Delta}}\left(\omega\right), \bar{\vec{\Delta}}\left(\omega-\Omega\right), \cdots]$. In the second equality, we keep terms up to the second order in $\kappa$, and then truncate $\underline{\hat{D}}$ to a $3\times3$ Floquet matrix(note that each matrix element of a Floquet matrix here is also a $4\times4$ matrix in the Keldysh$\otimes$Nambu space).  As will see in the following content, the transition temperature will be modified in this case, but not in the $\mathcal{O}(\kappa)$ case discussed in Sec.\ref{Sec:SPA}. The retarded part of one matrix element of the Floquet matrix $\underline{\hat{D}}$, defined as 
\begin{equation}
\hat{D}^{cl,q}(n,\omega)=-i\langle\Delta^{cl}(\omega+n\Omega)\bar{\Delta}^{q}(\omega)\rangle
\end{equation}
following from Eq.\eqref{Eq:BGF} is enough for the following discussion. The $(n,\omega)$ component of the Floquet matrix---$\underline{\hat{D}}$ has the following physical meaning: a bosonic excitation $\Delta(\omega)$ originally has energy $\omega$; due to the external driven field, it will absorb $n$ parts of energy($n\Omega$), and then becomes a bosonic mode with energy $\omega+n\Omega$. If $\hat{D}^{cl,q}(n,\omega)$ has a pole, then this process is inevitable, as poles of Green's functions correspond to quasiparticle excitations. Owing to structures of the distribution function and Green's functions $\breve{Q}_{\mathbf{k}}(t,t^{\prime})$, no poles exist when $\omega\neq0$. At $\omega=0$, this bosonic excitation is similar to the equilibrium counterpart. But, due to the  periodic driving, two pairing electrons can form bosonic modes with energy $n\Omega$, and generate the SC order parameter at higher harmonics. Then, the order parameter becomes periodic in time and has higher harmonics $\Delta^q(t)=\sum_{n}\Delta^q_ne^{i n \Omega t}$ (i.e. the {\em ansatz} we made in Sec.\ref{Sec:SPA}). In order to confirm this structure, we still need to check that if the bosonic modes at different harmonics occur at the same phase transition.

\begin{figure}[t]
\centering
\includegraphics[height=4.2cm]{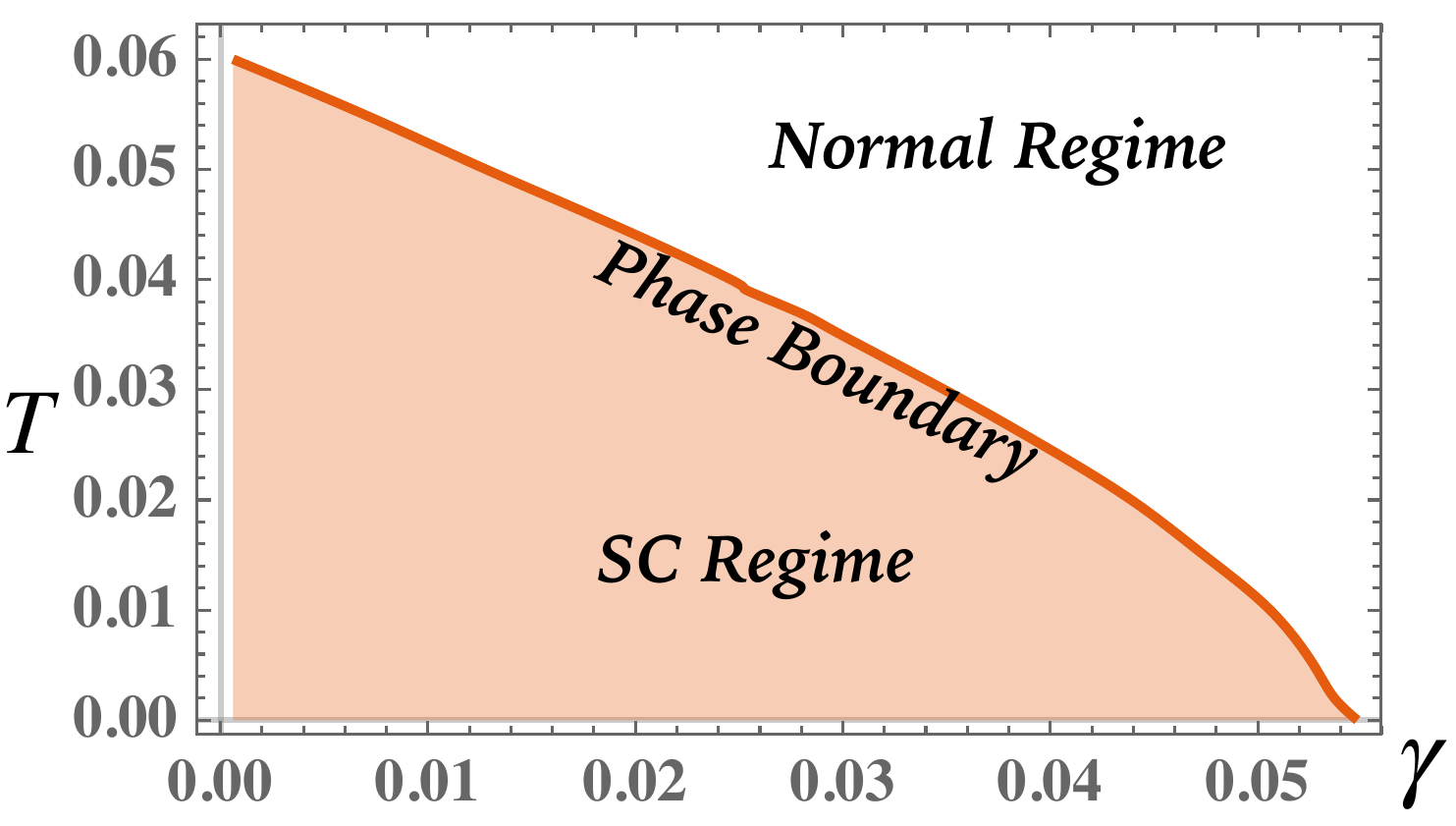}
\caption{Phase Diagram for $(\gamma,T)$. We choose $\omega_D/\Omega=10,\Omega=100K$, $g\rho_F=0.2$ and $\kappa=0.4$. $T$ and $\gamma$ are rescaled by being divided by $\Omega$.}
\label{Fig:PD}
\end{figure}

\begin{figure*}[t]
\centering
\includegraphics[height=5.2cm]{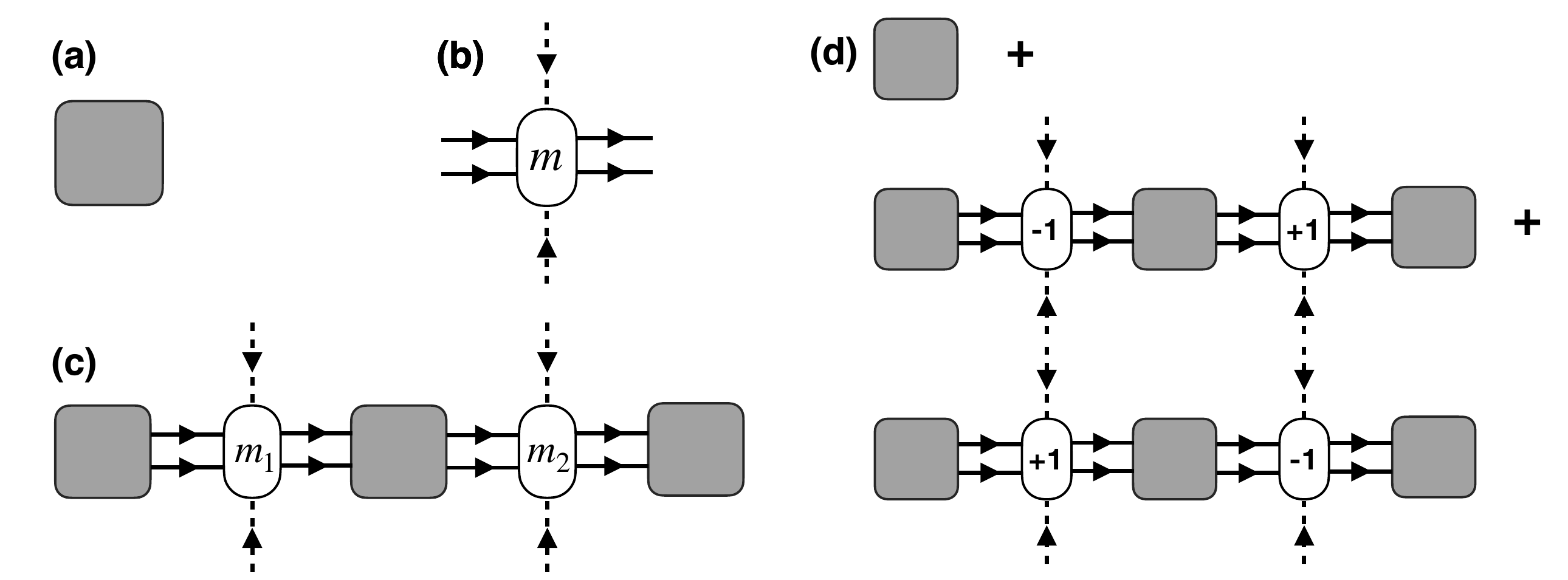}
\caption{Diagrams for different vertices. (a) Direct vertex; (b) Energy exchange vertex; (c) Indirect vertex; (d) Diagram for $-\hat{D}^{cl,q}(0,\omega)$.}
\label{Fig:DEIV}
\end{figure*}

We first consider the $n=0$ component---$\hat{D}^{cl,q}(0,0)$, which will result in zero-energy modes as the equilibrium case. The pole equation for the bosonic Green's function(see Appendix \ref{App:BGF}) can be written as 
\begin{equation}\label{pe}
1-g\sum_{\mathbf{k}}\hat{\Gamma}^{cl,q}_{\mathbf{k}}(0,0)=0,
\end{equation}
one will find the solution at zero temperature
\begin{equation}\label{netc}
\gamma=\gamma_c\equiv\omega_D^{\frac{1}{1-\frac{\kappa^2}{2}}}e^{-\frac{1+\frac{\kappa^2}{2}g\rho_F\ln\Omega}{g\rho_F(1-\frac{\kappa^2}{2})}}.
\end{equation}
which holds for $\gamma\ll\Omega\ll\omega_D$ and terms are kept up to $\mathcal{O}(\kappa^2)$ for deriving this. This result reduces to the SC transition temperature in equilibrium case for $\kappa\rightarrow 0$. For finite temperature, we numerically plot the phase diagram in Fig.\ref{Fig:PD}.
Note that $\hat{D}^{cl,q}(0,0)$ can also be interpreted as the normalized attractive interaction constant, and thus the divergence suggests that two electrons with opposite momenta and spins will form a bound state, known as Cooper pair. Therefore, we can see that our dissipative Floquet system could still develop a clear fermion-to-boson phase transition with a modification in the transition temperature due to the driving field. 

We now analyze Green's functions $\hat{D}^{cl,q}(\pm n,0)$ for $n>0$, which involves at most $n$ order transition processes between different Floquet bands. For example, after absorbing $m$ parts of energies($m\Omega$), the bosonic mode will transist from $a\Omega$ to $(a+m)\Omega$ with $|m|\leq n$. In that sense, we require perturbation calculations to keep terms up to $\mathcal{O}(\kappa^n)$. 

Here, we develop diagram rules to facilitate this analysis. For simplicity, we will introduce diagram rules through an example. 

When we study Cooper instability in the equilibrium case, the most important factor is the vertex of the two-electron correlation function, and under the random phase approximation(RPA), there is just one kind of vertices~\cite{Altland}. However, in our Floquet system, there will be more kinds of vertices, as particles can absorb or emit energies.

Comparing to the equilibrium case, one can observe that $-\hat{D}^{cl,q}(n,\omega)$ is the so-called vertex. As for $-\hat{D}^{cl,q}(0,0)$ shown in Appendix \ref{App:BGF}, it has three terms and they represent different kinds of vertices. Briefly, all vertices can be classified into two classes. One represents the direct process without emitting or absorbing energies, known as the {\em direct vertex}, and the other describes the indirect process containing energy exchanges, known as the {\em indirect vertex}. For example, in $-\hat{D}^{cl,q}(0,0)$, the first term represents the direct vertex, and the last two terms give indirect vertices.

As for direct vertices, they are just the same with those in the equilibrium case~\cite{Altland}. However, for indirect vertices, we should take absorbing and emitting processes into consideration, which only appears in the periodically driven system. They are described by $\hat{\Gamma}^{cl,q}(m,\omega)$ and we call them the {\em energy exchange vertices}. The index $m$ denotes the number of energies absorbed or emitted by particles(the energy unit is $\Omega$), and they absorb energy if $m>0$ and emit energy if $m<0$. $\omega$ denotes the sum of initial  energies of the two  scattering electrons. Through the analytical expression, one can find that indirect vertices are constructed from direct vertices and energy exchange vertices.  As it should be, the total number of energy exchanges should be consistent with $n$ in $-\hat{D}^{cl,q}(n,\omega)$.

Following above descriptions, we now write down diagram rules for different vertices, which will significantly facilitate the calculation of $\hat{D}^{cl,q}(n,\omega)$:
\begin{enumerate}
\item Attach $g/[1-g\hat{\Gamma}^{cl,q}(0,\omega)]$ to a direct vertex shown in Fig.\ref{Fig:DEIV}(a);
\item Attach $\hat{\Gamma}^{cl,q}(m,\omega)(\sim \kappa^{|m|}+\mathcal{O}(\kappa^{|m|+2}))$ to an energy exchange vertex shown in Fig.\ref{Fig:DEIV}(b);
\item The indirect vertex shown in Fig.\ref{Fig:DEIV}(c) is constructed from direct vertices and indirect vertices, thus one should attach 
\begin{equation*}
\begin{aligned}
&\qquad\frac{g}{1-g\hat{\Gamma}^{cl,q}(0,\omega)}\hat{\Gamma}^{cl,q}(m_1,\omega)\frac{g}{1-g\hat{\Gamma}^{cl,q}(0,\omega+m_1\Omega)}\\
&\qquad\times\hat{\Gamma}^{cl,q}(m_1,\omega+m_2\Omega)\frac{g}{1-g\hat{\Gamma}^{cl,q}(0,\omega+(m_1+m_2)\Omega)}
\end{aligned}
\end{equation*} 
to an indirect vertex. Read from the left to the right.
\item As for the diagram of one complete process, we just need to sum up those relative direct vertices and indirect vertices. As an example, $-\hat{D}^{cl,q}(0,0)$ can be expressed as Fig.\ref{Fig:DEIV}(d) with $\omega=0$, when we keep terms up to $\mathcal{O}(\kappa^2)$.
\end{enumerate}

Having shown the diagram rules for Floquet vertices or the bosonic Green's functions(In Appendix \ref{App:BGF}, we also summarize the diagram rules and show more examples for clarity.), one can now turn to analyze $\hat{D}^{cl,q}(n,0)$ used for demonstrating the periodicity of the bosonic condensation. 

As we now want to study the general case, that is $n$ can be arbitrary non-zero integers, terms should be kept up to $\mathcal{O}(\kappa^n)$. Therefore, suppose that we keep terms up to $\mathcal{O}(g^2\kappa^n)$ for simplicity. According to diagram rules discussed previously, we can observe that the number of direct vertices are no morn than $2$, as one direct vertex will contribute a factor $g$. Thus, there is one energy exchange vertex at most. For $\hat{D}^{cl,q}(\pm n,0)$ with $n\neq0$, there must be energy exchange vertices. Therefore, there is one energy exchange vertex and two direct vertices for the diagram of $-\hat{D}^{cl,q}(\pm n,0)$ in $\mathcal{O}(g^2\kappa^n)$ case. Thus, it is easy to obtain
\begin{equation}
\hat{D}^{cl,q}(\pm n,0)=\frac{-g^2\hat{\Gamma}^{cl,q}(\pm n,0)}{\left[1-g\hat{\Gamma}^{cl,q}(0,\pm n\Omega)\right]\left[1-g\hat{\Gamma}^{cl,q}(0,0)\right]},
\end{equation}
where we have $\hat{\Gamma}^{cl,q}(\pm n,\omega)\equiv\sum_{\mathbf{k}}\hat{\Gamma}_{\mathbf{k}}^{cl,q}(\pm n,\omega)$. Since $\hat{\Gamma}^{cl,q}(\pm n,\omega)\sim\kappa^n+O(\kappa^{n+2})$, $\hat{D}^{cl,q}(\pm n,0)\sim g^2\kappa^n$. In the $\mathcal{O}(\kappa^n)$ case, terms that is proportional to $\kappa^{|m|}$ with $|m|<n$ are still under our consideration, and one can  derive similar expressions for $\hat{D}^{cl,q}(m,0)$ from the same procedure. Note that the pole of $\hat{D}^{cl,q}(\pm n,0)$ is the same as that of $\hat{D}^{cl,q}(0,0)$ shown in Eq. \eqref{pe}. We also note that the term $1-g\hat{\Gamma}^{cl,q}(0,\pm n\Omega)$ in the denominator is always non-zero. For arbitrary higher order corrections of interaction constant $g$, the term $1-g\hat{\Gamma}^{cl,q}(0,0)$ always appears in their denominator as $\hat{D}^{cl,q}(\pm n,0)$. Thus, all Green's functions---$\hat{D}^{cl,q}(\pm n,0)$ exhibit the same pole structure (effective transition temperature $\gamma_c$) for all different $n$'s. The discussion above suggests that the bosonic excitation is periodic in the time domain, and can be expanded as $\Delta(t)=\sum_{n}\Delta(n\Omega)e^{in\Omega t}$. Due to the presence of the pole structure, the bosonic modes in the condensation are propagating and dissipationless even with system-bath coupling. This result also confirms the {\em ansatz} we made in Sec.\ref{Sec:SPA}. 




\section{Discussion and Summary}\label{Con}
The DOS of quasiparticles in Fig.\ref{Fig:DOS&Deltat}(b) shows that in the finite $\gamma$, there still exists energy levels in the gap. Actually, this comes from the fact that the lifetime of quasiparticles are now finite and they are diffusive, which can be derived from the Green's function of quasiparticles---$\breve{Q}_{\mathbf{k},\Delta}^R(0,\omega)$, Eq.\eqref{eq:GFoFQ}. Just like the broadening of the peak in the DOS of dissipative Floquet Majorana  
zero modes shown in Ref.\cite{yang2020dissipative}. One can think this is because in the presence of the bath, the fermionic degrees of freedom or electrons before the SC phase transition are diffusive, which can be found from the Green's functions Eq.\eqref{eq:GFoE}. However, the surprising thing is the bosonic mode or condensation  resulting from the superconducting phase transition is still a propagating mode in spite of the existence of dissipations, as the Green's functions of these bosonic modes---$\hat{D}^{cl,q}(\pm n,0)$, do not contain imaginary parts in the denominator.

In summary, based on the functional Keldysh field theory with a self-consistent treatment of all building blocks of our system, we demonstrate that the BCS mean-field treatment in the dissipative Floquet case is equivalent to the stationary point analysis in the functional or path integral formalism, where one can always implement the stationary point analysis. Moreover, based on the Gaussian fluctuation approximation, which is beyond the mean-field theory, we also consolidate the validity of the BCS mean-field theory in the dissipative Floquet scenario. Note that this system possesses a more structured gauge $U(1)$ symmetry, which could be an interesting point to be discussed in the future.




\begin{acknowledgments} 
D.E.L thanks Roman Lutchyn and Alex Levchenko for the inspired discussions to form the initial motivation of the project. The work is supported by National Science Foundation of China (Grant No. NSFC-11974198, Grant No. NSFC11888101), and the startup grant from State Key Laboratory of Low-Dimensional Quantum Physics and Tsinghua University.
\end{acknowledgments}

\begin{appendix}
\begin{widetext}
\section{Obtaining the Bosonic Effective Action}\label{app:EffAction}
In this section, we show some key steps that lead to the effective action Eq.\eqref{eq:BosonicAction}.

We start from the total Hamiltonian in the rotating frame, shown in Eq.\eqref{eq:RotatingFrame}
\begin{equation}
\begin{split}
H_F(t)=\sum_{\mathbf{k}\sigma}(\epsilon_{\mathbf{k}}-\mu_0)c^{\dagger}_{\mathbf{k}\sigma}c_{\mathbf{k}\sigma}+H_{int}+W\sum_{\mathbf{k}\mathbf{q}\sigma}\left(e^{if(t)}c^{\dagger}_{\mathbf{k}\sigma}a_{\mathbf{q}\sigma}+h.c.\right)+H_B, 
\end{split}\end{equation}
where expressions of $H_{int}$ and $H_B$ can be found in the main text, below Eq.\eqref{eq:OriginalHamiltonian}.

The next step is to write the action from this Hamiltonian, and the procedure is quite standard. For completeness, we start from the construction of the functional Keldysh field theory(see Ref.\cite{Kamenev} for detals). In the rotating frame, our system is governed by the Hamiltonian $H_F(t)$. The evolution of the density matrix can expressed as $\rho(t)=U_{t,-\infty} \rho(-\infty)\left(U_{t,-\infty}\right)^{\dagger}$, where $U_{t, t^{\prime}}=\mathbb{T} \exp \left(-i \int_{t^{\prime}}^{t} d t \hat{H}_F(t)\right)$ is the evolution operator, and $\mathbb{T}$ is the time-ordering operator. The expectation value of some observable can be calculated through the {\em generating function}, defined as 
\begin{equation}\label{eq:GF}
Z[V] \equiv \frac{\operatorname{Tr}\left[U_{C}[V] \rho(-\infty)\right]}{\operatorname{Tr}[\rho(-\infty)]},
\end{equation}
where $U_{C}[V] \equiv U_{-\infty,+\infty}[V] U_{+\infty,-\infty}[V]$, and $U_{\cdots}[V]$ is generated by $\hat{H}_{V}^{\pm}(t) \equiv \hat{H}_F(t) \pm \mathcal{O} V(t)$, where the plus(minus) sign refers to the forward(backward) part of the contour, Fig.\ref{Fig:CTCF}(a). $V(t)$ is an auxiliary field, and will be set to zero after taking the derivative of $Z[V]$ with respect to $V(t)$. 

\begin{figure}[t]
	\centerline{\includegraphics[height=4cm]{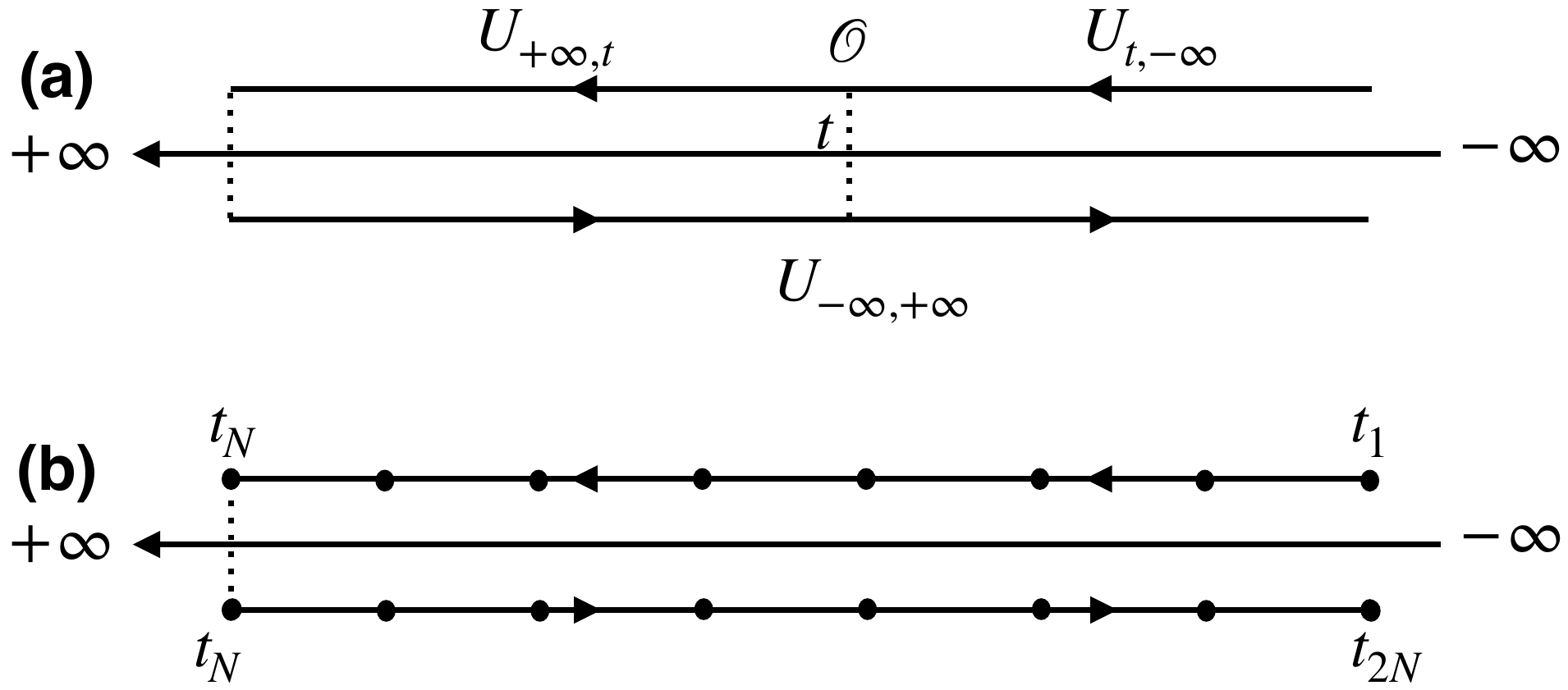}}
	\caption{(a) The closed time contour formalism is needed for a time-dependent system; (b) Discretize the closed time contour to develop the functional formalism. 
		\label{Fig:CTCF}}
\end{figure}
Eq.\eqref{eq:GF} implies that calculating the generating function is the key for deal with a many-body problem. The functional formalism(path integral formalism) is a useful method to rewrite the generating function. The standard procedure, which can be found in any quantum field theory textbook, is: 
\begin{itemize}
\item Divide the closed time contour into $(2N-2)$ intervals with length $\delta_t$, Fig.\ref{Fig:CTCF}(b);
\item Insert the revolution of unity in the coherent state basis(for fermions, we need the Grassmann number, denoted as $\psi$ in the following, for help);
\item Take the limit $\delta_t\rightarrow 0$, and then one will get the generating function $Z=\int D[\bar{\Psi},\Psi]\operatorname{exp}(iS)$ in the continuum limit. For our system, we have  
\begin{equation}
\begin{aligned}
S &=\int_cdt\int_cdt^{\prime}\sum_{\mathbf{k}}\vec{\Psi}^{\dagger}_{s\mathbf{k}}(t)\hat{Q}_{s0\mathbf{k}}^{-1}(t-t^{\prime})\vec{\Psi}_{s\mathbf{k}}(t^{\prime})+S_{int}\\
&\quad+\int_cdt\int_cdt^{\prime}\sum_{\mathbf{q}}\vec{\Psi}^{\dagger}_{b\mathbf{q}}(t)\hat{Q}_{b0\mathbf{q}}^{-1}(t-t^{\prime})\vec{\Psi}_{b\mathbf{q}}(t^{\prime})\\ 
&\quad+\int_cdt\sum_{\mathbf{{kq}}}\left[\vec{\Psi}^{\dagger}_{s\mathbf{k}}(t)\hat{M}(t)\vec{\Psi}_{b\mathbf{q}}(t)+h.c.\right].
\end{aligned}
\end{equation}
The definition of those quantities can be found in the main text. Here, we do not include $V(t)$, as we will not use it. 
\end{itemize}

Integrating out the degrees of freedom of the bath using Gaussian integrals, one can obtain the equivalent fermionic action:
\begin{equation}
S^{\prime}=\int_cdt\int_cdt^{\prime}\sum_{\mathbf{k}}\vec{\Psi}^{\dagger}_{s\mathbf{k}}(t)\hat{Q}_{\mathbf{k}}^{-1}(t,t^{\prime})\vec{\Psi}_{s\mathbf{k}}(t^{\prime})+S_{int},
\end{equation}
where
\begin{equation}\label{eq:DGF2}
\begin{aligned}
\hat{Q}_{\mathbf{k}}(t,t^{\prime})&=\hat{Q}_{s0\mathbf{k}}(t-t^{\prime})+\int_{-\infty}^{+\infty}dt_1dt_2\hat{Q}_{s0\mathbf{k}}(t-t_1)\hat{\Sigma}_{\mathbf{k}}(t_1,t_2)\hat{Q}_{\mathbf{k}}(t_2,t^{\prime}),
\end{aligned}
\end{equation}
with the self-energy from the bath being $\hat{\Sigma}_{\mathbf{k}}(t_1,t_2)=\sum_{\mathbf{q}}\hat{M}(t_1)\hat{Q}_{b0\mathbf{q}}(t_1-t_2)\hat{M}^{\dagger}(t_2)$, and 
\begin{equation*}
\hat{M}(t)=\left[
\begin{array}{ccc}
We^{if(t)} & 0\\
0 & -We^{-if(t)}
\end{array}
\right],
\end{equation*}
where $f(t)=(K/\Omega)\sin(\Omega t)$.
We now turn to discuss how to calculate the dressed Green's function $\hat{Q}_{\mathbf{k}}(t,t^{\prime})$ through the perturbative expansion with respect to $\kappa=K/\Omega$.  \\

Using the Fourier transformation Eq.\eqref{Eq:FT}, Eq.\eqref{eq:DGF2} can be rewritten as 
\begin{equation}\label{eq:DEiFS}
\hat{Q}_{\mathbf{k}}(n,\omega)=\delta_{n0}\hat{Q}_{s0\mathbf{k}}+\sum_{n_1}\hat{Q}_{s0\mathbf{k}}(\omega+n\Omega)\hat{\Sigma}_{\mathbf{k}}[n_1,\omega+(n-n_1)\Omega]\hat{Q}_{\mathbf{k}}(n-n_1,\omega),
\end{equation}
where $\hat{\Sigma}_{\mathbf{k}}(n,\omega)=\sum_{\mathbf{q}}\sum_{n_2}\hat{M}_{n+n_2}\hat{Q}_{b0\mathbf{q}}(\omega-n_2\Omega)\hat{M}_{n_2}^{\dagger}$, and 
\begin{equation}
\hat{M}_n=\left[\begin{array}{cc}W J_{n}\left(\frac{K}{\Omega}\right) & 0 \\ 0 & -W(-1)^{n} J_{n}\left(\frac{K}{\Omega}\right)\end{array}\right],
\end{equation}
where $J_n(x)$ is the Bessel function of the first kind. For weak driving amplitude cas, i.e., $\kappa\ll 1$, one can expand the Bessel function to the lowest order in $\kappa$. Up to $\mathcal{O}(\kappa^2)$, we have 
\begin{equation}
\hat{M}_{0}=\left[\begin{array}{cc}W & 0 \\ 0 & -W\end{array}\right],\; \hat{M}_{1}=\left[\begin{array}{cc}\frac{W}{2} \kappa & 0 \\ 0 & \frac{W}{2} \kappa\end{array}\right],\; \hat{M}_{-1}=\left[\begin{array}{cc}-\frac{W}{2}\kappa & 0 \\ 0 & -\frac{W}{2} \kappa\end{array}\right].
\end{equation}
Then, according to the Dyson equation Eq.\eqref{eq:DEiFS},  we can derive the dressed Green's function in the frequency space.

Next, we need to deal with the interaction action $S_{int}$, which is four-fermion interaction. One can treat this term through the Hubbard-Stratonovich transformation~\cite{Kamenev}. The idea of this transformation is that we multiply the generating function $Z$ by the unity
\begin{equation}
\mathbf{1}=\int D[\bar{\Delta},\Delta]\;\operatorname{exp}\left[i\int_cdt\;(-\frac{1}{g})\bar{\Delta}(t)\Delta(t)\right],
\end{equation} 
where $\Delta$ is a complex bosonic field. Thus, we have
\begin{equation}
e^{iS_{int}}=\int D[\bar{\Delta},\Delta]\;\operatorname{exp}\left[i\int_cdt\;g\sum_{\mathbf{k}\mathbf{k}^{\prime}}\bar{\psi}_{\mathbf{k}\uparrow}\bar{\psi}_{-\mathbf{k}\downarrow}\psi_{-\mathbf{k}^{\prime}\downarrow}\psi_{\mathbf{k}^{\prime}\uparrow}-\frac{1}{g}\bar{\Delta}\Delta\right],
\end{equation}
then we make a variable shift
\begin{equation}
\begin{aligned}
\bar{\Delta}&\rightarrow\bar{\Delta}-g\sum_{\mathbf{k}}\bar{\psi}_{\mathbf{k}\uparrow}\bar{\psi}_{-\mathbf{k}\downarrow},\\
\Delta&\rightarrow \Delta-g\sum_{\mathbf{k}^{\prime}}\psi_{-\mathbf{k}^{\prime}\downarrow}\psi_{\mathbf{k}^{\prime}\uparrow},
\end{aligned}
\end{equation}
and $e^{iS_{int}}$ becomes
\begin{equation}\label{eSint}
e^{iS_{int}}=\int D[\bar{\Delta},\Delta]\;\operatorname{exp}\left\{-i\int_c dt\left[\frac{1}{g}\bar{\Delta}\Delta-\bar{\Delta}\sum_{\mathbf{k}^{\prime}}\psi_{-\mathbf{k}^{\prime}\downarrow}\psi_{\mathbf{k}^{\prime}\uparrow}-\Delta\sum_{\mathbf{k}}\bar{\psi}_{\mathbf{k}\uparrow}\bar{\psi}_{-\mathbf{k}\downarrow}\right]\right\}.
\end{equation}
One can find that the generating function $Z=\int D[\bar{\Delta},\Delta,\bar{\Psi},\Psi]\operatorname{exp}\left[iS(\bar{\Delta},\Delta,\bar{\Psi},\Psi)\right]$ now is in a quadratic form with respect to the fermionic field, which means we can also integrate them out using Gaussian integrals. Finally, one will obtain the effective bosonic action Eq.\eqref{eq:BosonicAction}.

\section{Solving the Stationary Point Equation}\label{App:SSPE}
The key to solve the stationary point equation is to derive $\breve{Q}_{\mathbf{k},\Delta}(n,\omega)$. Since $\underline{\breve{Q}_{\mathbf{k}}^{-1}}=\underline{\breve{Q}_{s0\mathbf{k}}^{-1}}-\underline{\breve{\Sigma}_{\mathbf{k}}}$(derived from the Dyson equation Eq.\eqref{Eq:DE}),
where $\underline{\breve{Q}_{s0\mathbf{k}}^{-1}}$ is the free fermionic Green function and $\underline{\breve{\Sigma}_{\mathbf{k}}}$ is the self-energy provided by the normal metal bath in the Keldysh$\otimes$Nambu space. Note that $\underline{\breve{Q}_{\mathbf{k},\Delta}}=\left(\underline{\breve{Q}_{\mathbf{k},\Delta}^{-1}}\right)^{-1}$, we have the following expansion:
\begin{equation}\begin{aligned}
\underline{\breve{Q}_{\mathbf{k},\Delta}}
&=\left[\underline{\breve{Q}_{\mathbf{k},\Delta}^{-1(0)}}-\left[\underline{\breve{\Sigma}_{\mathbf{k}}^{\left(1\right)}}+\frac{1}{\sqrt{2}}\left(\underline{\Delta^{\alpha\left(1\right)}}\otimes\left(\hat{\gamma}^{\bar{\alpha}}\otimes\hat{\tau}_{+}\right)+\underline{\bar{\Delta}^{\alpha\left(1\right)}}\otimes\left(\hat{\gamma}^{\bar{\alpha}}\otimes\hat{\tau}_{-}\right)\right)\right]+\mathcal{O}(\kappa^2)\right]^{-1}\\
&\equiv\underline{\breve{Q}_{\mathbf{k},\Delta}^{\left(0\right)}}+\underline{\breve{Q}_{\mathbf{k},\Delta}^{\left(1\right)}}+\mathcal{O}\left(\kappa^{2}\right),
\end{aligned}\end{equation}
where the superscript $(i)$ also stands for the $i$th order in $\kappa$,
\begin{equation}\begin{split}
\underline{\breve{Q}_{\mathbf{k},\Delta}^{-1(0)}}\equiv
\underline{\breve{Q}_{s0\mathbf{k}}^{-1}}-\underline{\breve{\Sigma}_{\mathbf{k}}^{\left(0\right)}}-\frac{1}{\sqrt{2}}\left[\underline{\Delta^{\alpha\left(0\right)}}\otimes\left(\hat{\gamma}^{\bar{\alpha}}\otimes\hat{\tau}_{+}\right)+\underline{\Delta^{\alpha\left(0\right)}}\otimes\left(\hat{\gamma}^{\bar{\alpha}}\otimes\hat{\tau}_{-}\right)\right],\label{Q0}
\end{split}\end{equation}
and
\begin{equation}\begin{aligned}\label{Q1kD}
\underline{\breve{Q}_{\mathbf{k},\Delta}^{(1)}}\equiv\underline{\breve{Q}^{(0)}_{\mathbf{k},\Delta}}\left[\underline{\breve{\Sigma}^{\left(1\right)}_{\mathbf{k}}}+\frac{1}{\sqrt{2}}\left(\underline{\Delta^{\alpha\left(1\right)}}\left(\hat{\gamma}^{\bar{\alpha}}\otimes\hat{\tau}_{+}\right)+\underline{\bar{\Delta}^{\alpha\left(1\right)}}\left(\hat{\gamma}^{\bar{\alpha}}\otimes\hat{\tau}_{-}\right)\right)\right]\underline{\breve{Q}^{(0)}_{\mathbf{k},\Delta}}.
\end{aligned}\end{equation}
Due to the feature of Floquet matrices, $\underline{\breve{Q}_{\mathbf{k},\Delta}^{(0)}}$ is a diagonal matrix, and $\underline{\breve{Q}_{\mathbf{k},\Delta}^{(1)}}$ is a secondary diagonal matrix, etc. 

For $\Delta^q_0$, we have 
\begin{equation}\begin{aligned}\label{ACd0q}
\Delta_{0}^{q}&=i\frac{g}{\sqrt{2}}\int\frac{d\omega}{2\pi}\operatorname{Tr}\left[(\hat{\gamma}^{q}\otimes\hat{\tau}_{-})\sum_{\mathbf{k}}\breve{Q}_{\mathbf{k},\Delta}\left(0,\omega\right)\right]\\
&=i\frac{g}{\sqrt{2}}\int\frac{d\omega}{2\pi}\sum_{\mathbf{k}}\left[\breve{Q}_{\mathbf{k},\Delta}^{K}\left(0,\omega\right)\right]^{12},
\end{aligned}\end{equation}
where the superscript ''$12$'' stands for the matrix element in the first row and second column. When we keep terms up to $\mathcal{O}(\kappa)$, normal fermionic Green's functions with dissipations are
\begin{equation}\begin{aligned}\label{eq:GFoE}
\breve{Q}_{\mathbf{k},p}^{R/A}\left(0,\omega\right)&=\frac{1}{\omega-\epsilon_{\mathbf{k}}\pm i\gamma},\quad
\breve{Q}_{\mathbf{k},h}^{R/A}\left(0,\omega\right)=\frac{1}{\omega+\epsilon_{\mathbf{k}}\pm i\gamma},\quad
\breve{Q}_{\mathbf{k},p/h}^{K}\left(0,\omega\right)=-2 i \gamma\frac{\tanh(\frac{\omega}{2T})}{\left(\omega\mp\epsilon_{\mathbf{k}}\right)^{2}+\gamma^{2}}.
\end{aligned}\end{equation}
From $\breve{Q}_{\mathbf{k},p/h}^K(0,\omega)$, we find that when we keep terms up to $\mathcal{O}(\kappa)$, the fermionic distribution function is still the Fermi-Dirac distribution, which is different from that in $\mathcal{O}(\kappa^2)$ case~\cite{Liu}. Then Eq.\eqref{Q0} leads to the Green's functions of quasiparticles, dressed by the bosonic field $\Delta$:
\begin{equation}\begin{aligned}\label{eq:GFoFQ}
\breve{Q}_{\mathbf{k},\Delta}^{R/A}\left(0,\omega\right)=\frac{1}{\left(\omega\pm i\gamma\right)^{2}-\epsilon_{\mathbf{k}}^{2}-\Delta^{2}}
\begin{bmatrix}\omega+\epsilon_{\mathbf{k}}\pm i\gamma & \Delta\\
\Delta & \omega-\epsilon_{\mathbf{k}}\pm i\gamma
\end{bmatrix},
\end{aligned}\end{equation}
and
\begin{equation}\begin{aligned}
\breve{Q}_{\mathbf{k}, \Delta}^{K}(0, \omega)=\breve{Q}_{\mathbf{k}, \Delta}^{R}(0, \omega)\cdot F(\omega)-F(\omega)\cdot \breve{Q}_{\mathbf{k}, \Delta}^{A}(0, \omega),
\end{aligned}\end{equation}
where we have assumed $\Delta$ to be real, $F(\omega)=\left[1-2n_F(\omega)\right]\mathbb{1}_{N}$ with $\mathbb{1}_{N}$ being the identity in the Nambu space and $n_F(\omega)=1/(e^{\beta \omega}+1)$ is the Fermi-Dirac distribution.

Substituting $\sum_{\mathbf{k}}$ with $\int d\epsilon_{\mathbf{k}}\rho_F$, one will get the quasi-classical Green's function~\cite{Liu,RevModPhys.88.035005,ch2003introduction} defined as $\breve{Q}_{\Delta}^K=\sum_{\mathbf{k}} \breve{Q}_{\mathbf{k},\Delta}^{K}$:
\begin{equation}\begin{aligned}
\quad\left[\breve{Q}_{\Delta}^{K}\left(0,\omega\right)\right]^{12}
=i\pi \rho_{F}\left[\frac{\Delta\tanh\frac{\omega}{2T}}{\sqrt{\left(\omega-i\gamma\right)^{2}-\Delta^{2}}}-\frac{\Delta\tanh\frac{\omega}{2T}}{\sqrt{\left(\omega+i\gamma\right)^{2}-\Delta^{2}}}\right].
\end{aligned}\end{equation}
Then according to Eq.\eqref{ACd0q}, one will get the gap equation in the main text.

Now we turn to consider $\Delta^q_{\pm1}$. Seeing as $\underline{\breve{Q}^{(1)}_{\mathbf{k},\Delta}}$ is not a diagonal matrix, for convenience, we truncate the Floquet matrix to a $3\times3$ one. Then, 
 one can solve the matrix equation Eq.\eqref{Q1kD} to get 
\begin{equation}
\breve{Q}_{\mathbf{k},\Delta}(\pm1,\omega)=\breve{Q}^{(0)}_{\mathbf{k},\Delta}(0,\omega\pm\Omega)
\left\{\breve{\Sigma}_{\mathbf{k}}(\pm1,\omega)+\frac{1}{\sqrt{2}}\left[\Delta^{\alpha}_{\pm1}(\hat{\gamma}^{\bar{\alpha}}\otimes\hat{\tau}_{+})+\bar{\Delta}^{\alpha}_{\pm1}(\hat{\gamma}^{\bar{\alpha}}\otimes\hat{\tau}_{-})\right]\right\}\breve{Q}^{(0)}_{\mathbf{k},\Delta}(0,\omega).
\end{equation}
Solving this equation, one can derive $\breve{Q}^K_{\mathbf{k},\Delta}(\pm1,\omega)$, then one can get equations that decide $\Delta^q_{\pm1}$. Since it is difficult to analytically solve them, we resort to numerical calculations.

\section{Bosonic Green's Functions and The Diagram Rules}\label{App:BGF}
In this section, we briefly discuss the derivation of the bosonic Green's functions. Here, we will keep terms up to the second order in $\kappa$. As shown in the main text, the transition temperature will be modified in this case, but not in the $\mathcal{O}(\kappa)$ case. The second order approximation in $\Delta^{\alpha}$($\bar{\Delta}^{\alpha}$) with $\alpha\in\{cl,q\}$ of the effective bosonic action---$S_{\operatorname{eff}}$ has been derived and shown in Eq.\eqref{Eq:SOAoBEA}. Then, according to the feature of Gaussian integrals, one can get 
\begin{equation}\begin{split}
\underline{\hat{D}}&=-i\langle\begin{bmatrix}\vec{\Delta}\left(\omega+\Omega\right)\\
\vec{\Delta}\left(\omega\right)\\
\vec{\Delta}\left(\omega-\Omega\right)
\end{bmatrix}
\begin{bmatrix}\vec{\bar{\Delta}}\left(\omega+\Omega\right) & \vec{\bar{\Delta}}\left(\omega\right) & \vec{\bar{\Delta}}\left(\omega-\Omega\right)
\end{bmatrix}\rangle\\
&=\begin{bmatrix}
-\frac{1}{g}\hat{\sigma}_{1}+\hat{\Gamma}\left(0,\omega+\Omega\right) & \hat{\Gamma}\left(1,\omega\right) & \hat{\Gamma}\left(2,\omega-\Omega\right)\\
\hat{\Gamma}\left(-1,\omega+\Omega\right) & -\frac{1}{g}\hat{\sigma}_{1}+\hat{\Gamma}\left(0,\omega\right) & \hat{\Gamma}\left(1,\omega-\Omega\right)\\
\hat{\Gamma}\left(-2,\omega+\Omega\right) & \hat{\Gamma}\left(-1,\omega\right) & -\frac{1}{g}\hat{\sigma}_{1}+\hat{\Gamma}\left(0,\omega-\Omega\right)
\end{bmatrix}^{-1}\\
&=-g\left(\underline{\mathbb{1}}-g\underline{\hat{\Gamma}^{(0)}}\otimes\hat{\sigma}_1-g\underline{\hat{\Gamma}^{(1)}}\otimes\hat{\sigma}_1-g\underline{\hat{\Gamma}^{(2)}}\otimes\hat{\sigma}_1\right)^{-1}\otimes\hat{\sigma}_{1}
\end{split}\end{equation}
where we have truncated the Floquet matrix to a $3\times3$ one, $\hat{\Gamma}\equiv\sum_{\mathbf{k}}\hat{\Gamma}_{\mathbf{k}}$, the superscript $(i)$ denotes terms kept up to the $i$th order in $\kappa$,
\begin{equation}
\underline{\mathbb{1}}-g\underline{\hat{\Gamma}^{(0)}}\otimes\hat{\sigma}_1=\left[
                                                                    \begin{array}{ccc}
                                                                      \mathbb{1}_K-g\hat{\Gamma}(0,\omega+\Omega)\hat{\sigma}_1 & 0 & 0 \\
                                                                      0 & \mathbb{1}_K-g\hat{\Gamma}(0,\omega)\hat{\sigma}_1 & 0 \\
                                                                      0 & 0 & \mathbb{1}_K-g\hat{\Gamma}(0,\omega-\Omega)\hat{\sigma}_1 \\
                                                                    \end{array}
                                                                  \right],
\end{equation}
\begin{equation}
 g\underline{\hat{\Gamma}^{(1)}}\otimes\hat{\sigma}_1=\left[
                                                                \begin{array}{ccc}
                                                                  0 & g\hat{\Gamma}(1,\omega)\hat{\sigma}_1 & 0 \\
                                                                  g\hat{\Gamma}(-1,\omega+\Omega)\hat{\sigma}_1 & 0 & g\hat{\Gamma}(1,\omega-\Omega)\hat{\sigma}_1 \\
                                                                  0 & g\hat{\Gamma}(-1,\omega)\hat{\sigma}_1 & 0 \\
                                                                \end{array}
                                                              \right],
\end{equation}
and
\begin{equation}
  g\underline{\hat{\Gamma}^{(2)}}\otimes\hat{\sigma}_1=\left[
                                                                \begin{array}{ccc}
                                                                  0 & 0 & g\hat{\Gamma}(2,\omega-\Omega)\hat{\sigma}_1\\
                                                                  0 & 0 & 0 \\
                                                                 g\hat{\Gamma}(-2,\omega+\Omega)\hat{\sigma}_1 & 0 & 0 \\
                                                                \end{array}
                                                              \right],
\end{equation}
where $\mathbb{1}_K$ is the identity matrix in the Keldysh space. Expanding $(\underline{\mathbb{1}}-g\underline{\hat{\Gamma}}\otimes\hat{\sigma}_1)^{-1}$ up to $\kappa^2$, one will get
\begin{equation}\label{Eq:D0w}
\begin{split}
&\quad\left(\underline{\mathbb{1}}-g\underline{\hat{\Gamma}}\otimes\hat{\sigma}_1\right)^{-1}\\
&\cong\left(\underline{\mathbb{1}}-g\underline{\hat{\Gamma}^{(0)}}\otimes\hat{\sigma}_1\right)^{-1}+\left(\underline{\mathbb{1}}-g\underline{\hat{\Gamma}^{(0)}}\otimes\hat{\sigma}_1\right)^{-1}\cdot g\underline{\hat{\Gamma}^{(1)}}\otimes\hat{\sigma}_1\cdot \left(\underline{\mathbb{1}}-g\underline{\hat{\Gamma}^{(0)}}\otimes\hat{\sigma}_1\right)^{-1}\\
&\quad+\left(\underline{\mathbb{1}}-g\underline{\hat{\Gamma}^{(0)}}\otimes\hat{\sigma}_1\right)^{-1}\cdot g\underline{\hat{\Gamma}^{(2)}}\otimes\hat{\sigma}_1\cdot\left(\underline{\mathbb{1}}-g\underline{\hat{\Gamma}^{(0)}}\otimes\hat{\sigma}_1\right)^{-1}\\
&\quad+\left(\underline{\mathbb{1}}-g\underline{\hat{\Gamma}^{(0)}}\otimes\hat{\sigma}_1\right)^{-1}\cdot g\underline{\hat{\Gamma}^{(1)}}\otimes\hat{\sigma}_1\cdot\left(\underline{\mathbb{1}}-g\underline{\hat{\Gamma}^{(0)}}\otimes\hat{\sigma}_1\right)^{-1}\cdot g\underline{\hat{\Gamma}^{(1)}}\otimes\hat{\sigma}_1\cdot\left(\underline{\mathbb{1}}-g\underline{\hat{\Gamma}^{(0)}}\otimes\hat{\sigma}_1\right)^{-1}+\mathcal{O}(\kappa^3).
\end{split}
\end{equation}
Then one can get matrix elements of $\hat{D}^{cl,q}(n,\omega)$ up to the second order in $\kappa$:
\begin{equation}
\begin{split}
\hat{D}^{cl,q}\left(0,\omega\right)&=-\frac{g}{1-g\hat{\Gamma}^{cl,q}\left(0,\omega\right)}\\
&\quad-\frac{g^{3}\hat{\Gamma}^{cl,q}\left(1,\omega-\Omega\right)\hat{\Gamma}^{cl,q}\left(-1,\omega\right)}{\left[1-g\hat{\Gamma}^{cl,q}\left(0,\omega\right)\right]\left[1-g\hat{\Gamma}^{cl,q}\left(0,\omega-\Omega\right)\right]\left[1-g\hat{\Gamma}^{cl,q}\left(0,\omega\right)\right]}\\
&\quad-\frac{g^{3}\hat{\Gamma}^{cl,q}\left[-1,\omega+\Omega\right)\hat{\Gamma}^{cl,q}\left(1,\omega\right]}{\left[1-g\hat{\Gamma}^{cl,q}\left(0,\omega\right)\right]\left[1-g\hat{\Gamma}^{cl,q}\left(0,\omega+\Omega\right)\right]\left[1-g\hat{\Gamma}^{cl,q}\left(0,\omega\right)\right]},\label{p1200}
\end{split}\end{equation}
\begin{equation}\label{Eq:D1}
\hat{D}^{cl,q}(\pm1,\omega)=-\frac{g^{2}\hat{\Gamma}^{cl,q}\left(\pm1,\omega\right)}{\left[1-g\hat{\Gamma}^{cl,q}\left(0,\omega\pm\Omega\right)\right]\left[1-g\hat{\Gamma}^{cl,q}\left(0,\omega\right)\right]},
\end{equation}
\begin{equation}\begin{split}\label{Eq:D2}
\hat{D}^{cl,q}\left(\pm2,\omega\right)&=-\frac{g^2\hat{\Gamma}^{cl,q}\left(\pm2,\omega\right)}{\left[1-g\hat{\Gamma}^{cl,q}\left(0,\omega\right)\right]\left[1-g\hat{\Gamma}^{cl,q}\left(0,\omega\pm2\Omega\right)\right]}\\
&\quad-\frac{g^{3}\hat{\Gamma}^{cl,q}\left(\pm1,\omega\right)\hat{\Gamma}^{cl,q}\left(\pm1,\omega\pm\Omega\right)}{\left[1-g\hat{\Gamma}^{cl,q}\left(0,\omega\right)\right]\left[1-g\hat{\Gamma}^{cl,q}\left(0,\omega\pm\Omega\right)\right]\left[1-g\hat{\Gamma}^{cl,q}\left(0,\omega\pm2\Omega\right)\right]}.
\end{split}
\end{equation}

In order to make the diagram rules we show in the main text clearer, we summarize them again and give one more example. 

The diagram rules read as:
\begin{enumerate}
\item Attach $g/[1-g\hat{\Gamma}^{cl,q}(0,\omega)]$ to a direct vertex shown in Fig.\ref{Fig:DEIV}(a);
\item Attach $\hat{\Gamma}^{cl,q}(m,\omega)(\sim \kappa^{|m|}+\mathcal{O}(\kappa^{|m|+2}))$ to an energy exchange vertex shown in Fig.\ref{Fig:DEIV}(b);
\item The indirect vertex shown in Fig.\ref{Fig:DEIV}(c) is constructed from direct vertices and indirect vertices, thus one should attach 
\begin{equation*}
\begin{aligned}
\qquad\frac{g}{1-g\hat{\Gamma}^{cl,q}(0,\omega)}\hat{\Gamma}^{cl,q}(m_1,\omega)\frac{g}{1-g\hat{\Gamma}^{cl,q}(0,\omega+m_1\Omega)}\hat{\Gamma}^{cl,q}(m_1,\omega+m_2\Omega)\frac{g}{1-g\hat{\Gamma}^{cl,q}(0,\omega+(m_1+m_2)\Omega)}
\end{aligned}
\end{equation*} 
to an indirect vertex. Read from the left to the right.
\item As for the diagram of one complete process, we just need to sum up those relative direct vertices and indirect vertices. 
\end{enumerate}

In the main text, we use the example of $\hat{D}^{cl,q}(0,0)$ to introduce the diagram rules, and in the following, we will use the diagram rules to derive $\hat{D}^{cl,q}(\pm2,\omega)$, and comparing it to Eq.\eqref{Eq:D2} to show the correctness of the diagram rules. Here, we also keep terms up to the second order in $\kappa$.

For $\hat{D}^{cl,q}(\pm2,\omega)$, since there exist energy exchange processes($n=\pm2$), thus in the diagram, the energy exchange vertex must exist. 
\begin{itemize}
\item The simplest process is that there are only one energy exchange vertex, which stands for one particle absorbing/emitting two parts of energies($\pm2\Omega$), and two direct vertices. The diagram is shown in Fig.\ref{Fig:DfD2}(a);

\item One may find that a process that contains two energy exchange vertices, which stands for one particle first absorbing/emitting one part of energy($\pm\Omega$), then absorbing/emitting one part of energy as well, can also contribute to $\hat{D}^{cl,q}(\pm2,\Omega)$. It indeed does, and the diagram is shown in Fig.\ref{Fig:DfD2}(b);

\item Since the leading order term of the process, in which one particle absorbs/emits $n$ parts of energies, is proportional to $\kappa^n$(that is $\hat{\Gamma}^{cl,q}(\pm n,\omega)\sim\kappa^n+\mathcal{O}(\kappa^{n+2})$), in the $\mathcal{O}(\kappa^2)$ case, there are only the above two diagrams that contribute to $\hat{D}^{cl,q}(\pm2,\omega)$. The diagram for $\hat{D}^{cl,q}(\pm2,\omega)$ is just the sum of Fig.\ref{Fig:DfD2}(a) and Fig.\ref{Fig:DfD2}(b).

\item Use the diagram rules shown previously, one can easily derive the expression of $\hat{D}^{cl,q}(\pm2,\omega)$:
\begin{equation}\begin{split}
-\hat{D}^{cl,q}\left(\pm2,\omega\right)&=\frac{g^2\hat{\Gamma}^{cl,q}\left(\pm2,\omega\right)}{\left[1-g\hat{\Gamma}^{cl,q}\left(0,\omega\right)\right]\left[1-g\hat{\Gamma}^{cl,q}\left(0,\omega\pm2\Omega\right)\right]}\\
&\quad+\frac{g^{3}\hat{\Gamma}^{cl,q}\left(\pm1,\omega\right)\hat{\Gamma}^{cl,q}\left(\pm1,\omega\pm\Omega\right)}{\left[1-g\hat{\Gamma}^{cl,q}\left(0,\omega\right)\right]\left[1-g\hat{\Gamma}^{cl,q}\left(0,\omega\pm\Omega\right)\right]\left[1-g\hat{\Gamma}^{cl,q}\left(0,\omega\pm2\Omega\right)\right]},
\end{split}
\end{equation}
which is exactly the same with Eq.\eqref{Eq:D2}.
\end{itemize}

\begin{figure}[t]
\centering
\includegraphics[height=3.5cm]{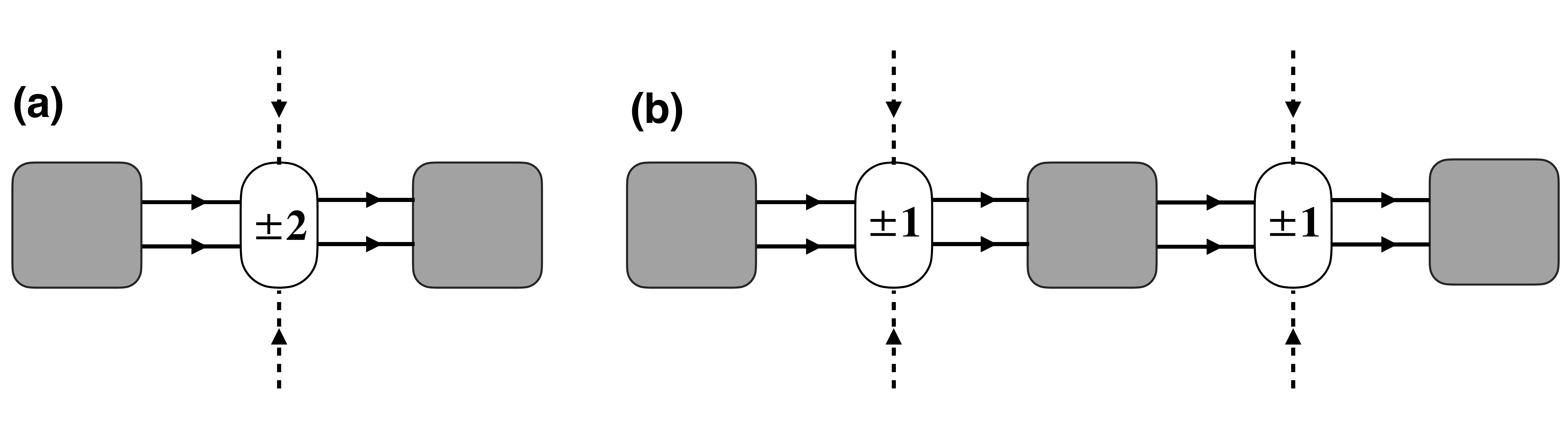}
\caption{Diagrams for $\hat{D}^{cl,q}(\pm2,\omega)$. (a) One particle absorbs/emits two parts of energies($\pm2\Omega$); (b) One particle first absorbs/emits one part of energy($\pm\Omega$), then absorbs/emits one part of energy as well.}
\label{Fig:DfD2}
\end{figure}

\end{widetext} 
\end{appendix}

\bibliography{references,FloquetK1,FloquetETH_OPEN,DissipativeMaj,FloquetMajoranas}
\bibliographystyle{apsrev4-1}

\end{document}